\documentclass[aps,pre,twocolumn,showpacs,superscriptaddress]{revtex4}
\usepackage{amsmath}
\usepackage{amsfonts}
\usepackage{amssymb}
\usepackage{epsfig}
\usepackage{color}
\newcommand{\be}{\begin{equation}}
\newcommand{\ee}{\end{equation}}

\begin{document}

\title{Inhomogeneous shear flows in soft jammed 
materials with tunable attractive forces}

\author{Pinaki Chaudhuri}
\affiliation{Laboratoire PMCN, 
Universit\'e Claude Bernard Lyon 1, Villeurbanne, France}

\author{Ludovic Berthier}
\affiliation{Laboratoire Charles Coulomb, 
UMR 5221, CNRS and Universit\'e Montpellier 2, Montpellier, France}

\author{Lyd\'eric Bocquet}
\affiliation{Laboratoire PMCN, 
Universit\'e Claude Bernard Lyon 1, Villeurbanne, France}

\date{\today}

\begin{abstract}
We perform molecular dynamics simulations to characterize the occurrence of 
inhomogeneous shear flows in soft jammed materials.
We use rough walls to impose a simple shear flow and study 
the athermal motion of jammed assemblies of soft particles, both 
for purely repulsive interactions and in the presence of an 
additional short-range attraction of varying strength.
In steady state, pronounced flow inhomogeneities emerge for all systems 
when the shear rate becomes small. Deviations 
from linear flow are stronger in magnitude
and become very long-lived when the strength of the
attraction increases, but differ from permanent shear-bands.
Flow inhomogeneities occur in a stress window
bounded by the dynamic and static yield stress values. 
Attractive forces enhance the flow heterogeneities because they
accelerate stress relaxation, thus effectively moving the system closer 
to the yield stress regime where inhomogeneities are most pronounced.
The present scenario for understanding the effect of particle adhesion
on shear localization, which is based on detailed molecular dynamics 
simulations with realistic particle interactions, differs 
qualitatively from previous qualitative explanations 
and ad-hoc theoretical modelling. 
\end{abstract}

\pacs{62.20.-x, 83.60.La, 83.80.Iz}


\maketitle

\section{Introduction}

Soft jammed materials, such as dense emulsions, foams and 
pastes are ubiquitous in nature and have a 
wide range of industrial applications~\cite{vanhecke,coussot2}.
Normally, these materials only flow
when an externally applied stress exceeds a critical value, the 
``yield stress'', while they behave as a soft solid otherwise. 
Thus, the flow properties of these systems are intrinsically non-linear and 
exhibit complex features that challenge both experimentalists
and theoreticians~\cite{coussot}. 
To understand the rheology of these complex fluids, we want to know the 
mechanical response of the system to an externally applied force. 
However, this becomes a highly non-trivial task when the response
is not spatially homogeneous, or when it occurs 
over a broad range of timescales. These two challenging 
factors routinely characterize the rheology of soft jammed
materials, emphasizing the need for spatially and temporally
resolved studies of the flow properties in dense particulate systems
under shear~\cite{vanhecke2010}. Clearly, molecular dynamics simulations 
are well-suited to pursue this task, because they naturally combine 
particle resolution with the possibility to simulate 
large systems with controlled interactions 
over relatively long times.
 
In typical experimental conditions, a broad variety of complex
fluids display spatially inhomogeneous flows, usually described as
``shear-bands'', even though this single name in fact hides a diversity
of distinct phenomena~\cite{vanhecke2010}. In this paper, we are specifically
interested in the flow behaviour 
of dense assemblies of large spherical particles that form athermal disordered
solids, and we primarily think of foams, emulsions and dense colloidal 
suspensions as relevant experimental realizations of our 
numerical model~\cite{vanhecke2010}.
While early experiments commonly reported the existence of inhomogeneous
shear-banded flows in foams and dense emulsions, more recent 
work~\cite{cohen,manneville,becu} 
has established that, when properly prepared and studied 
over sufficiently long times, shear-bands in dense systems of soft 
repulsive particles do not appear as a permanent phenomenon, although
flow inhomogeneities may appear to be extremely long-lived
in some cases~\cite{manneville}. Recently, however, it was 
reported in several instances that the addition 
of a small amount of attractive forces between particles 
triggers the appearance of 
shear-bands~\cite{becu,ragouillaux,ovarlez,paredes}. 
A common interpretation
is that shear-bands in this case result from thixotropic behaviour 
competing with the imposed flow~\cite{bonnclass}. While natural
for low-density colloidal gels which have a complex 
structure~\cite{gels,denn}, 
this explanation appears less convincing for jammed systems, which present
instead a fairly homogeneous structure. 

Shear-banding phenomena have been observed also in numerical simulations of 
amorphous solids under flow~\cite{jlanael,book}. 
It was initially reported for a sheared Lennard-Jones mixture~\cite{varnik},
where it was argued that the coexistence of flowing and static 
phases results from the existence of
distinct bounds for static and dynamic yield stresses
leading to a multivalued flow curve~\cite{berthier03}. Subsequently,
in studies of model metallic glasses~\cite{shi}, shear-banding was observed 
using a variety of boundary conditions, quench rates, 
or systems sizes, which motivated theoretical extensions of 
the shear-transformation zone model~\cite{stz} 
to account for shear-bands~\cite{manning}.
However, in experiments performed on actual metallic glasses, these flow 
inhomogeneities evolve rapidly with the applied deformation 
and the system develops fractures before a steady state can be reached.
In a parallel effort, studies of athermal quasi-static shear 
flow of amorphous solids have revealed 
the existence of system-spanning avalanches generated by correlated
activation of plastic events~\cite{lemaitremaloney,tanguy}. These 
observations seem to be in tune with the generic scenario that 
dynamical heterogeneities are a characteristic feature of amorphous 
materials~\cite{book}. Although it is tempting 
to speculate that dynamic heterogeneities, avalanches, 
and shear-bands are various facets of the same underlying physics,
more precise links between these phenomena are missing~\cite{jlanael}.

At the theoretical level, many early models developed to account for
the rheology of soft amorphous materials were mean-field 
in nature, and spatial fluctuations were usually 
discarded~\cite{sgr,hb,BBK}. More recently, several 
coarse-grained models have emerged that attempt to capture
the idea, revealed by the above mentioned numerical studies, 
that plastic events are localized but may trigger  
additional plastic events elsewhere in the system, thus 
cascading into sustained flow~\cite{picard,vdb,jagla,kirsten1,bocquet}. 
While such modelling 
directly yields spatially inhomogeneous dynamics, the appearance
of permanent shear-bands does not necessarily follow. 
Numerical simulations
of these models indeed do not produce genuine 
shear-bands~\cite{picard,vdb}, which seems to suggest
that shear-bands might only occur under quite specific conditions. 
Several recent studies of simple models suggest that some form of long-lived 
shear-banding phenomena may occur after shear 
start-up~\cite{manning,Moorcroft,roux2}. These approaches also build
on the possibility to observe distinct static and dynamic 
yield stress values, the former being enhanced by prolonged 
aging in thermal glasses.

Several mechanisms have been put forward to account for permanent
flow inhomogeneities, which typically revolve 
around the idea that the stress-strain rate flow curve, 
$\sigma = \sigma(\dot{\gamma})$, 
is multi-valued. We already mentioned the possibility, 
first discussed in \cite{berthier03,varnik}, that the 
flow curve at finite shear rates is monotonic [for instance
of the Herschel-Bulkley type with a finite dynamic 
yield stress, $\sigma_d = \lim_{\dot{\gamma} \to 0} 
\sigma(\dot{\gamma}) $], but that there exists a static 
branch at $\dot{\gamma} = 0$ extending up to a static yield stress
value, $\sigma_s$, 
larger than the dynamic one, $\sigma_s > \sigma_d$. 
This opens a finite range of stress 
where the shear rate can take two values. 
Genuine non-monotonic flow curves have recently been obtained 
in various models by including the generic idea that yielding dynamics 
should be self-consistently connected to the evolution of the 
local structure.
This was done for instance using self-consistent dynamics of the effective
temperature~\cite{fielding} in the 
Soft Glassy Rheology model~\cite{sgr,sollich},
or by incorporating thixotropic effects using 
an additional timescale for structural ``restructuration''
in schematic~\cite{coussotovarlez} or coarse-grained 
models~\cite{vincent,kirsten2,jagla}, in which case non-monotonic 
flow curves arise when structural recovery is slower than
the characteristic relaxation time of the system. 
This hypothesis was however not justified by microscopic
arguments or detailed measurements. Alternatively, 
it has been proposed  that shear-banding
could also occur due to a shear-concentration 
coupling~\cite{besseling}, with the possibility that at small
enough shear rates, variations in local concentrations would result in the 
large fluctuations in flow-rates. 
Clearly, more numerical and experimental studies are 
now necessary in order to test and discriminate 
these different physical ideas. 

In this paper, we report simulational studies of the athermal flow of 
highly jammed systems, consisting of particles 
having either purely repulsive interactions (as in foams~\cite{cohen} and 
simple emulsions~\cite{becu,ragouillaux}), 
or having both repulsive and short-range 
attractions (as in adhesive 
emulsions~\cite{becu,ragouillaux,ovarlez,paredes}). 
Motivated by the 
phenomonological finding that attractive forces might be responsible 
for the appearance of permanent shear-bands, we specifically 
check whether changing interactions
results in different shear localization properties during steady state flow,
and use our spatially and temporally resolved simulations 
to seek a physical interpretation of our findings. 
We find that strong flow inhomogeneities 
are present in all systems, except that the lifetime and degree of 
fluctuations are indeed much higher with increased attractions, but 
we do not observe simple, permanent shear-bands even in our most 
adhesive systems. We find instead that attractive forces, 
by accelerating stress relaxation, effectively shift the 
system closer towards the yield stress regime where 
flow inhomogeneities can become so pronounced that the system
is not able to sustain a linear flow profile. Such a theoretical scenario was 
was not anticipated in any of the above-mentioned work.

The paper is organized as follows. In Sec.~\ref{model} we 
describe our model and numerical methods. We then present
our measurements and results in Sec.~\ref{results}, and 
we discuss our results in Sec.~\ref{discussion}. We conclude
the paper in Sec.~\ref{conclusion}. 

\section{Models and numerical methods}
\label{model}

\subsection{Repulsive interactions} 

The model system that we study is a collection of polydisperse soft 
particles, introduced as a model for foams~\cite{durian,ohernfoam},
and which has now been extensively studied to understand the physics 
of jammed soft materials~\cite{vanhecke} both in 
athermal conditions~\cite{ohern1,ohern2} and at finite 
temperatures~\cite{witten}.

In the repulsive case, two particles, 
having diameters $d_i$ and $d_j$, interact 
via a harmonic potential:
\begin{equation}
\label{repulsion}
V(r_{ij}) =
\begin{cases}
\epsilon (1-\frac{r_{ij}}{d_{ij}})^2, & \text{$r_{ij} < d_{ij}$}\\
0, & \text{$r_{ij} > d_{ij}$}\\
\end{cases}
\end{equation}
where $d_{ij}=(d_i+d_j)/2$.
We choose a 50:50 binary mixture for the polydispersity with a mean 
diameter of $\left\langle d \right\rangle = 1.0$ 
and a size ratio of $1.4$ to avoid crystallization.

\subsection{Attractive interactions} 

\begin{figure}
\psfig{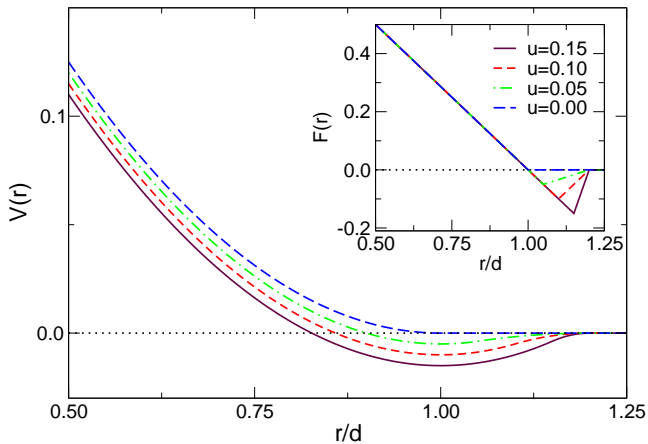}
\caption{Interparticle potential $V(r)$, Eqs.~(\ref{repulsion}, 
\ref{attractioneq}), 
for $\epsilon=1/2$, and tunable strength of attractive 
forces from $u=0$ (purely repulsive case)
to $u=0.15$. The range of the attractive forces
is constant, $r/d = 1.2$, but the strength increases with $u$.
Inset: The corresponding forces (shown with the same 
colours) with the labelling for $u=-\ell_1/d$ quantifying the strength 
of the attractive part.}
\label{fig1}
\end{figure}

We introduce adhesive interactions between the particles in 
a manner similar to models of cohesive granular media~\cite{ohernattract}.
Specifically, we introduce two parameters,
$\ell_1$ and $\ell_2 > \ell_1$, through which we can control 
the range and depth of the attractive forces between the particles.
We choose the following
form for the interparticle interactions:
\begin{equation}
\label{attractioneq}
V(r_{ij})=
\begin{cases}
\epsilon[(1-\frac{r_{ij}}{d_{ij}})^2-\ell_1 \ell_2],  
& \frac{r_{ij}}{d_{ij}} < 1+\ell_1 \\
\frac{-\epsilon \ell_1}{\ell_2 -\ell_1} 
[1+\ell_2-\frac{r_{ij}}{d_{ij}}]^2, 
& 1+\ell_1 < \frac{r_{ij}}{d_{ij}} < 1+\ell_2 \\
0, & \frac{r_{ij}}{d_{ij}} > 1+\ell_2 \\
\end{cases}
\end{equation}
This simple form is chosen because it yields an interaction 
force which is piecewise linear, see Fig.~\ref{fig1}.

In our simulations, we use $\ell_2$ to fix the range of the 
interparticle force: $\ell_2=0.2$,
and change the depth and range of the attractive part 
by varying $\ell_1$. The shape of the interactions, 
with the inclusion of the attractive part, is illustrated in 
Fig.~\ref{fig1}. By varying the position 
at distance $(1+\ell_1)d_{ij}$
of the minimum in the interparticle force, the amplitude 
of the attractive force is varied, which we quantified 
by introducing the parameter
$u=-2 \epsilon \ell_1/d_{ij}$,
which sets an energy scale for the attraction strength.
We have studied different values of the particle adhesion,
$u=0$, 0.05, 0.10, and 0.15, as shown in Fig.~\ref{fig1}. 
We do not explore larger adhesion strength to avoid the occurrence
of any shear-induced phase separation.

\subsection{Simulation methods} 

We study the shear flow of a two dimensional system of soft particles. 
We use very large dimensions for the simulation box, 
$L_x =84.46 \langle d \rangle$ and $L_y=99.39\langle d \rangle $, 
which contains $N=10404$ particles. 
Shear is imposed via rough walls constructed as follows. A layer of particles 
having thickness $2 \langle d \rangle$ is frozen both at the top and bottom 
of the simulation box in the $y$ direction, from an unsheared relaxed 
configuration. These rough walls have thus a structure similar 
to the sheared system, and the same two walls are used
throughout this work. In the simulations with attractive forces,
we use the same structure for the walls, but implement the 
same attractive forces for the wall-fluid interactions.

To impose a constant shear rate $\dot{\gamma}$, we pull the top
wall at a velocity fixed by $v_{\rm wall}=\dot{\gamma}(L_y-4 \langle d 
\rangle)$, 
with the bottom wall being kept fixed.
We also carry out some shear 
simulations with an imposed constant stress, $\sigma$. This is
done by pulling the top wall by a tangential force 
$F=\sigma L_x$~\cite{varnikyield,xu}, with the bottom wall
again remaining fixed. 

The motion of the particles
in the bulk are governed by the conservative forces described above, 
while athermal behaviour is ensured using viscous dissipative forces.
During the flow, when two particles overlap, they experience a 
dissipative force which depends on their relative velocity:
$-b [(\vec{v}_i-\vec{v}_j).\hat{r}_{ij}] \hat{r}_{ij} $, 
where $b=2$ is the damping coefficient, and 
$\hat{r}_{ij}$ is the unit vector between particles 
$i$ and $j$.
The range of the ``overlap'' used for dissipation is 
$d_{ij}$ for pairs of purely repulsive particles, 
and corresponds to 
$(1+\ell_2)d_{ij}$ when attractive forces are included.

We work at a constant volume fraction of $\phi= N \pi \left\langle 
d^3 \right\rangle / (6V) = 1.0$, 
which is much beyond the jamming point $\phi_J \sim 0.85$.
This implies that the structure is fairly homogeneous
and resembles the one of dense amorphous solids. Thus, 
our modelling approach bears no similarities with colloidal gels, 
which are found to also exhibit shear-bands in experimental 
work~\cite{gels,denn}. 

The units for energy, length and time are $2 \epsilon$, 
$\langle d \rangle$ and $\langle d \rangle/ 
\sqrt{ 2 \epsilon/ m}$, respectively, where 
$m$ is the mass of the particles.
The trajectories of the particles are evolved
by numerically integrating the corresponding Newton's equations of motion, 
using a velocity-Verlet scheme~\cite{degroot}.

\section{Measurements and Results}

\label{results}

\subsection{Repulsive Particles}

As discussed above, several recent experiments have suggested that jammed
materials made of soft repulsive particles do not exhibit steady state
shear-bands. For the system of harmonic spheres introduced by 
Durian to study wet foams~\cite{durian}, simulations have both 
reported the presence~\cite{xu} or absence~\cite{langlois}
of shear localization. The same model also displays
strong dynamical heterogeneities under shear, similar to unsheared 
thermal glasses, quantified
by measuring the usual dynamical susceptibility $\chi_4$~\cite{claus}. 
Such heterogeneities have also been recently measured in the flow of 
NIPAM particles which similarly 
interact solely via repulsive interactions~\cite{nordstorm}.
Additionally, in the quasi-static limit the same model 
displays system-spanning avalanches~\cite{claus2}
and strongly non-affine particle motion~\cite{claus3}.
Therefore, there are marked spatiotemporal fluctuations
and non-affine particle motion for this system under shear flow, 
which certainly need to be taken into 
account when discussing velocity profiles and their fluctuations.

\begin{figure}
\psfig{file=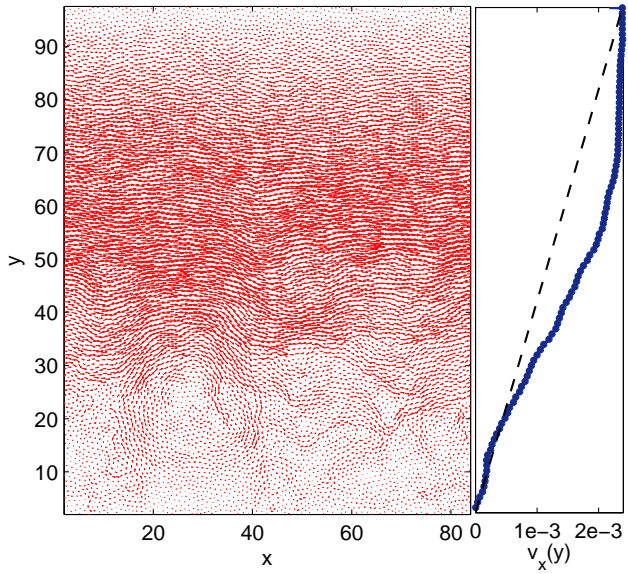,width=8.5cm}
\caption{Left: Map of non-affine displacements for repulsive particles, 
measured for an imposed shear-rate of $\dot{\gamma}=2.5\times{10^{-5}}$
and during a strain window of $\Delta\gamma=0.10$ taken during 
steady state flow. Right: The corresponding velocity profile, 
measured during the same strain window. The dashed line shows the velocity 
profile expected for a linear flow. A clear ``band'' can be observed
near the middle of the system, which spans the system horizontally, 
but it is not permanent.} 
\label{fignonaff1}
\end{figure}

We begin to explore kinetic heterogeneities by looking at the map of 
non-affine 
displacements  of the particles,
defined as deviations from the local single-particle displacements 
expected from assuming a linear velocity profile.
In the left panel of Fig.~\ref{fignonaff1}, we show such a map corresponding 
to a measurement done within a strain window
of $\Delta\gamma=0.10$ during steady state flow at an imposed strain rate of 
$\dot{\gamma} = 2.5 \times {10^{-5}}$, while the right panel
shows the corresponding velocity profile.
One can clearly observe the spatial heterogeneities
in the dynamics with regions having different mobilities within this period 
of deformation. 
Moreover, it is evident that the particles which have undergone large 
displacements cluster together to 
form a ``band'' aligned in the flow direction, 
similar to what has been for instance observed in amorphous
Lennard-Jones solids under quasistatic deformation~\cite{tanguy}, at 
finite shear rates at $T=0$~\cite{lemaitre}, or in supercooled
liquids~\cite{tanaka}. Quite often, the appearance of such 
system-spanning heterogeneities are invoked as proof of 
presence of ``shear-bands''.

\begin{figure}
\psfig{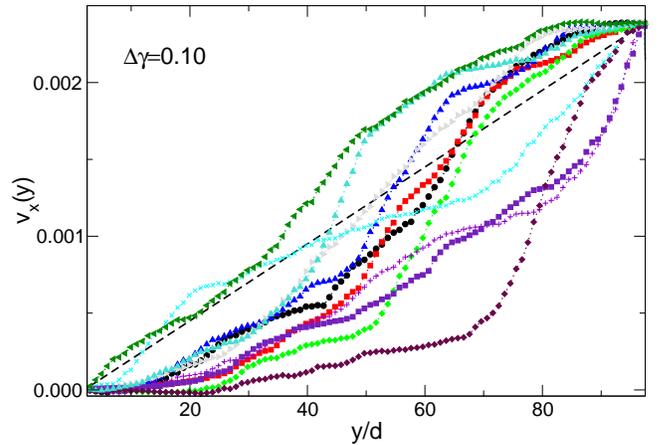}
\caption{A series of representative velocity profiles 
for repulsive particles, averaged during a strain window 
of $\Delta\gamma=0.10$, and sampled during
steady-state flow at an imposed $\dot{\gamma} = 2.5\times{10^{-5}}$.
These profiles reveal strong deviations from linear profiles, 
which strongly fluctuate both in space and time.}
\label{figvelprof1}
\end{figure}

The spatial variation of mobilities, observed during this strain window 
$\Delta{\gamma}$, should also be
reflected in the velocity profiles measured during the same interval. 
Indeed it does. The velocity profile
measured during the same strain window when the above map was generated 
is shown in the right panel of Fig.~\ref{fignonaff1}.
The region of large mobility in the map corresponds to a large deviation 
from a linear velocity profile.

A number of velocity profiles, all measured during 
independent strain windows of 
$\Delta{\gamma}=0.10$ during steady state flow at the same
value of the shear rate are shown in Fig.~\ref{figvelprof1}. 
Large fluctuations 
deviating away from the expected linear velocity profile
are generically observed. However, during 
flow the location of the more mobile region 
is seen to switch from one place to another, and the 
observed ``shear-bands'' are in fact not permanent but have their own 
dynamics. Moreover, we do not always see a clear ``coexistence''
between two distinct regions of 
mobility, which is often associated with 
shear-banding. Instead, the velocities might sometimes evolve 
more gradually across the channel.

The observation of velocity profiles suggests that 
before drawing conclusions about the presence of shear localization
it is necessary to study and quantify more precisely the degree
of fluctuations of the velocity profiles, in order
to answer the following two questions. How do the 
observed heterogeneities depend on the strain window
chosen to average the profile? How do these fluctuations 
depend on the imposed macroscopic shear rate?
We feel that such quantitative information is mandatory 
when reporting on shear-banding phenomena.

To analyze these fluctuations, we average the velocities in the 
$x$-direction to compute the local 
strain rates $\dot{\gamma}(y)$ averaged over a given 
strain window $\Delta{\gamma}$. One 
can then construct, for any chosen $\Delta{\gamma}$,
a histogram of these locally observed strain rates, 
which we denote $N(\dot{\gamma})$.
Clearly this probability distribution function depends
on the two key parameters whose influence we wish to study,
namely the strain window, $\Delta{\gamma}$,
and the macroscopically imposed shear rate, $\dot{\gamma}$. 

\begin{figure}
\psfig{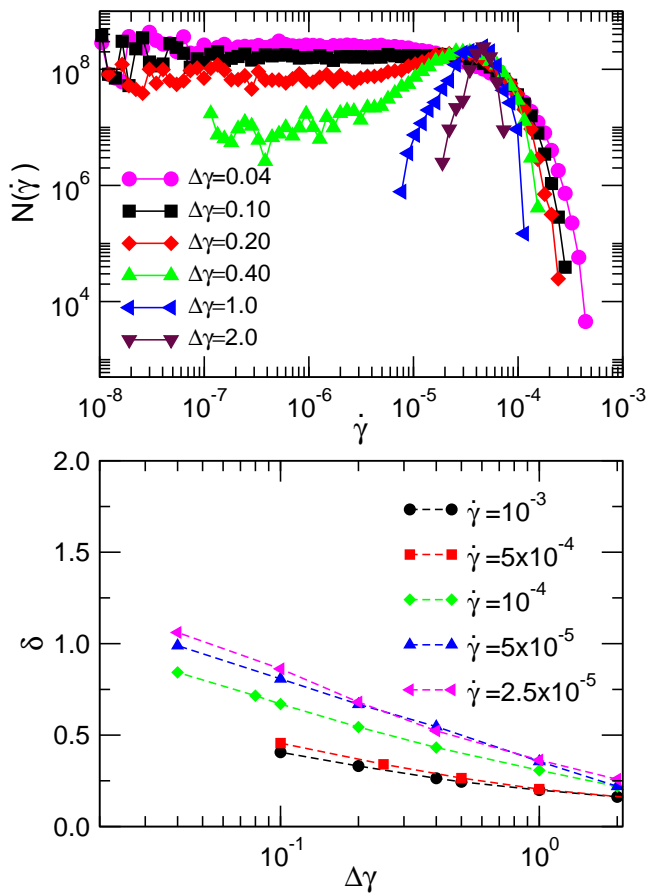}
\caption{
Top: Distribution of local strain rates $\dot{\gamma}(y)$
with changing the strain window $\Delta\gamma$ used for the average,
at an imposed strain rate of $5\times{10^{-5}}$.
Bottom: Variation of $\delta$, Eq.~(\ref{delta}),
which quantifies the spread of the distribution $N(\dot{\gamma})$, 
with the strain window $\Delta\gamma$, for a range of imposed strain rates.}
\label{repfig2}
\end{figure}

We first show the evolution of $N(\dot{\gamma})$ with 
the strain window for a fixed value of the shear rate,
$\dot{\gamma} = 5 \times{10}^{-5}$ in the top panel 
of Fig.~\ref{repfig2}. For small strain intervals,
$\Delta{\gamma}=0.04 - 0.10$,
$N(\dot{\gamma})$ spans across a wide range of local strain rates 
from nearly immobile regions 
($\dot{\gamma}\sim{10^{-8}}$) to regions which flow faster 
($\dot{\gamma}\sim{3\times{10^{-4}}}$) than
the imposed shear rate. Moreover, the shape of the distribution
is clearly not symmetric and non-Gaussian, with the 
appearance of a flat tail towards small values of the shear 
rate. This behaviour is clearly consistent with our observation
that spatial fluctuations in the profiles, shown in Fig.~\ref{figvelprof1},
actually span the entire system. However, the distribution narrows down and 
becomes closer to a Gaussian with an 
increase of the strain window 
over which velocity profiles are averaged, suggesting that 
spatial fluctuations become less correlated over time.
For the largest deformation, $\Delta{\gamma}=2$, the 
fluctuations around the imposed shear rate have become quite 
small. Our observations are similar to what was 
reported for a sheared Lennard 
Jones glass~\cite{onuki}. For small times, only a few plastic events occur 
resulting in the
initial large heterogeneities which are localized in space. However, if 
one waits long enough,
the plastic events proliferate across the system and the heterogeneities 
therefore are erased, which eventually results in homogeneous flow.
Intriguingly, such description 
of the transient character of shear inhomogeneities 
is also reminiscent of the temporal evolution of kinetic heterogeneities
characterizing the structural relaxation of thermal glassy 
systems~\cite{book}.

These results suggest that some form of ``shear-banding'' exists 
in the present system, but flow localization is not a permanent phenomenon.
It is thus natural to ask about the lifetime of these inhomogeneities.
To this end, we introduce the ``dispersity'' 
$\delta$ of the distribution $N(\dot{\gamma})$, as the ratio of the 
standard deviation to the mean of the distribution:
\begin{equation}
\delta = \frac{ \sqrt{ \left\langle \dot{\gamma}^2 \right\rangle_{N} 
- \langle 
\dot{\gamma} \rangle_{N}^2}
  } { \langle 
\dot{\gamma} \rangle_{N} }, 
\label{delta}
\end{equation}
where the average $\langle \cdots \rangle_N$ 
is performed over the probability distribution
of the local shear rate, $N(\dot{\gamma}$).
The dispersity $\delta$ is the most natural way to quantify the width
of this distribution. 

In the bottom panel of Fig.~\ref{repfig2}, we show the variation of $\delta$ 
with the strain interval
$\Delta{\gamma}$ for a range of imposed shear rates from 
$\dot{\gamma} = 2.5 \times 10^{-5}$ to $10^{-3}$. Following from our
discussion above, we find that $\delta$ decreases with $\Delta{\gamma}$ 
for all $\dot{\gamma}$. Therefore, if one averages the velocities 
for long enough strain windows, say larger than $\Delta{\gamma}\sim{2}$, 
nearly linear velocity profiles 
will be observed in all cases. In fact, it is interesting to note 
that for all values of $\dot{\gamma}$, heterogeneities 
become negligible at approximately the same strain interval $\Delta\gamma$.
However, in the regime of smaller 
$\Delta{\gamma}$, we see that flows become 
more heterogeneous with decreasing shear rates.

From these results, we conclude that our model of a jammed system
with soft repulsive interactions does not produce
permanent shear localization, although strong 
flow inhomogeneities are detected when insufficient averaging
of the velocity profiles is performed, which might be 
a relevant issue in  experiments. 
This analysis also suggests that 
a discussion of shear-banding in soft jammed materials 
can not be separated from a discussion of their spatio-temporal 
dynamics. In particular, we have presented in 
Fig.~\ref{repfig2} a simple method to quantify the lifetime
of these inhomogeneities. In Sec.~\ref{discussion} we will relate these
observations to the global rheology of the system.
 
\subsection{Including attractive interactions}

Motivated by recent experimental results~\cite{ragouillaux,becu} where the 
inclusion of particle adhesion in dense emulsions 
resulted in qualitatively different flow patterns 
compared to repulsive emulsions, we proceeded to 
explore the nature of flow heterogeneities 
in sheared soft disks with the tunable attractive interactions
shown in Fig.~\ref{fig1}. 

\begin{figure}
\psfig{file=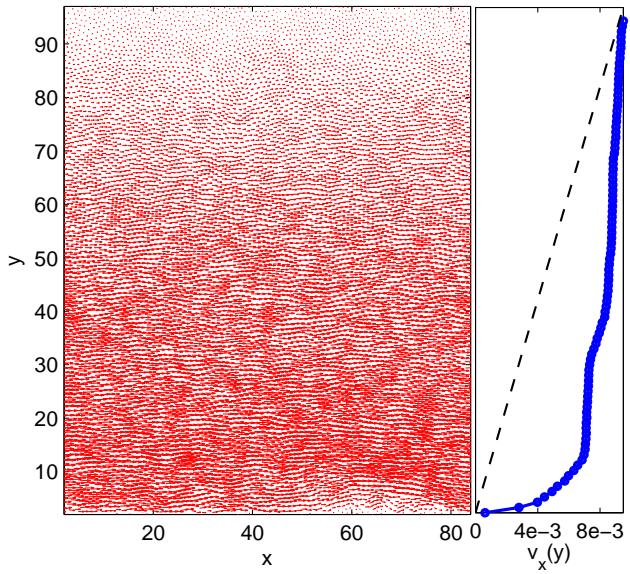,width=8.5cm}
\caption{Left: Map of non-affine displacements for attractive particles,
$u=0.15$,  measured for an imposed shear-rate 
of $\dot{\gamma}= 10^{-4}$
and during a strain window of $\Delta\gamma=0.10$ taken during 
steady state flow. Right: The corresponding velocity profile, 
measured during the same strain window. The dashed line shows the velocity 
profile expected for a linear flow. Compared with 
Fig.~\ref{fignonaff1}, the flow is much more inhomogeneous,
the top 80~\%  of the system being nearly unsheared, while the bottom
part near the wall is sheared very strongly.
\label{u15nonaffmap}}
\end{figure}

For these sticky particles, we again look at the map of non-affine 
displacements. In the left panel of Fig.~\ref{u15nonaffmap}, we show 
the spatial map 
of such displacements for $u=0.15$ during a strain interval of 
$\Delta\gamma=0.10$, measured in steady state at an imposed shear-rate of 
$\dot{\gamma}=10^{-4}$. 
The corresponding velocity profile is shown in the right panel of
Fig.~\ref{u15nonaffmap}.
In the bottom of the map, we can again very clearly 
see a ``band'' formed by the
most mobile particles, 
which spans across the entire length of the system in the flow direction,
and having a transverse width of around $10-20$ diameters,
while the top of the system is mainly unsheared. The 
corresponding velocity profile also reflects this via its strongly
nonlinear shape in the entire system. More
interestingly, the mobile band populates the bottom of the shear 
cell adjacent to the static wall, whereas the quiet ones are adjacent to the
top wall of the cell via which shear is generated across the cell, which 
is also evident from the corresponding velocty profile. 

\begin{figure}
\psfig{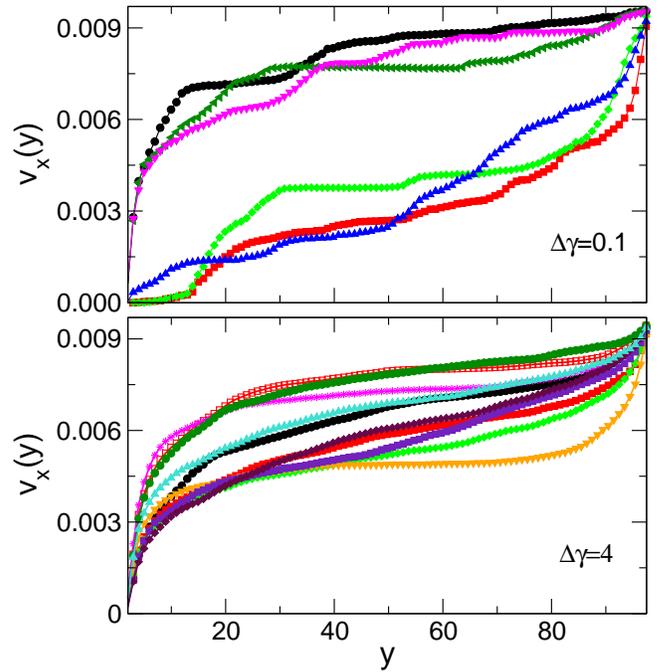}
\caption{A series of representative velocity profiles 
for attractive particles with $u=0.15$, averaged during a strain window 
of $\Delta\gamma=0.10$ (top), and $\Delta\gamma=4.0$ (bottom),
and sampled during steady-state flow at an 
imposed shear rate $\dot{\gamma} = 10^{-4}$.
These profiles reveal strong deviations from linear profiles, 
which strongly fluctuate both in space and time. Compared to the 
repulsive system, these profiles remain strongly non-linear 
at large deformation $\Delta \gamma=4.0$, suggesting that a linear 
velocity profile is not stable 
in the presence of strong particle 
adhesion. 
\label{attvelprofs}}
\end{figure}

This is further illustrated by looking at 
a set of consecutive velocity profiles 
measured during the same strain window, $\Delta\gamma=0.10$, as shown in 
the top panel of Fig.~\ref{attvelprofs}. We observe quite 
dramatic deviations from a 
linear velocity profile in all cases and one can clearly 
see a switching of the position of the more 
mobile population with time as the shear-band continuously flips from one 
wall to the other. Note that the shear rate at the center of the channel 
is always very small, such that
the bulk of the system either flows with the 
right wall, or remains immobile with the left wall. 
Although seemingly reminiscent of the velocity oscillations 
reported for colloidal particles in microchannels~\cite{isa}, 
the motion we observe with attractive particles is instead
very far from periodic, and it is enhanced by low 
rather than fast shear rates.

When averaged over a much longer
period, $\Delta\gamma=4$,
the velocity profiles still deviate significantly from linear profiles, 
as shown in the bottom panel of Fig.~\ref{attvelprofs}.
For the profiles measured during this period, the central region 
appears nearly unsheared, whereas the regions close to both the top
and bottom walls show local shear rates which are larger than 
the imposed value.
Thus, the flow heterogeneities are much more pronounced in the 
attractive system,  with much clearer signs of 
the ``coexistence'' between sheared
and unsheared regions, and this inhomogeneity 
seems to be persistent over much longer strain intervals 
for the attractive systems as compared to the repulsive one.

Are these ``shear-bands'' permanent objects? To answer 
this question for these sticky particles, we 
again characterize their lifetime, 
repeating the exercise performed for the repulsive particles. 
For each strength of attraction $u$, we 
compute local strain rates $\dot{\gamma}$ from velocity profiles 
averaged over different strain intervals $\Delta\gamma$.
We build the corresponding histograms, $N(\dot\gamma)$, and measure the
dispersity, $\delta$, of the local strain rates
from Eq.~(\ref{delta}). 
Since the lifetime of the inhomogeneities becomes large when 
attraction increases, it becomes numerically difficult to sample
a large number of independent fluctuations and get accurate 
statistics for the distributions. 
This forces us to impose a larger shear rate, 
$\dot{\gamma}=10^{-4}$, because much longer
simulations would be needed to obtain good statistics
at lower shear rates, where, presumably, even longer-lived 
and more pronounced are present.

\begin{figure}
\psfig{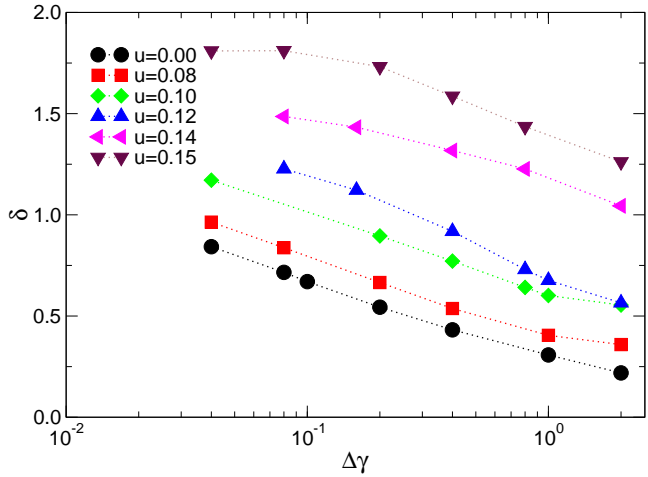}
\caption{Effect of particle attraction on the 
variation of the dispersity $\delta$, Eq.~(\ref{delta}),
which quantifies the spread of the distribution $N(\dot{\gamma})$, 
with the strain window $\Delta\gamma$, 
for an imposed shear rate of $\dot{\gamma}=10^{-4}$.
Attraction enhances both the amplitude and the lifetime 
of the flow heterogeneities.
\label{momdat1e4}}
\end{figure}

The data for $\dot{\gamma} = 10^{-4}$ are shown 
in Fig.~\ref{momdat1e4}. We can distinctly see that
for all strain intervals, increasing attraction between the particles
results in increased heterogeneities. 
We suspect the effect would be even more pronounced for lower shear 
rates. In fact, within the strain 
windows that we have been able to sample, the
heterogeneities have not died out when $u$ is large enough. For instance,
the dispersity obtained over an averaging window
$\Delta \gamma = 2.0$ for $u > 0.12$ is larger than the dispersity
for $u=0$ and $\Delta \gamma = 0.1$.  However, $\delta$
is still clearly a decreasing function of $\Delta \gamma$, which 
suggests that for even larger strain windows, the temporal 
fluctuations reported in Fig.~\ref{attvelprofs} 
eventually decrease the overall width of the shear rate
distribution $N(\dot{\gamma})$.
While we conclude that there is no
permament shear-bands in our jammed adhesive systems, 
our data are nevertheless in broad agreement with 
experimental results showing
that particle adhesion strongly enhance 
localization of 
the shear~\cite{becu,ragouillaux,paredes,ovarlez},
in the sense that our most adhesive systems do not sustain 
stable linear flow profiles, even when averaging 
velocities over deformations as large as 400~\%.

In the following section, we shall discuss the physical origin
of these observations.

\section{Discussion and interpretation}

\label{discussion}

\subsection{Static and dynamic yield stresses}

To understand the above observations and the effect of the 
attractive forces, we now turn to the global flow curve of the system
and ask which (if any) 
of the theoretical arguments summarized in the introduction
might apply to our system.

\begin{figure}
\psfig{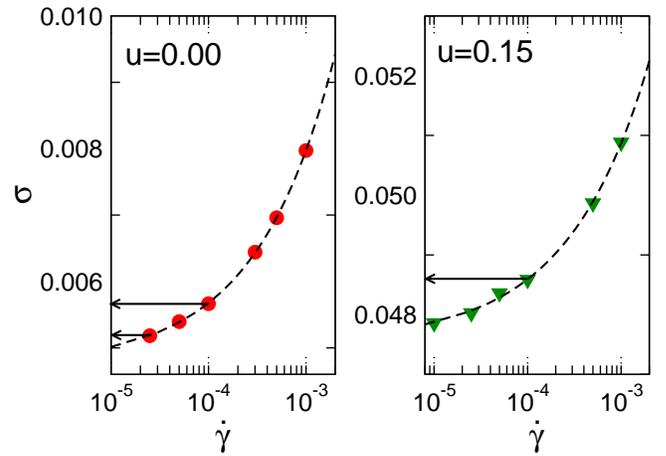}
\caption{Global flow curves $\sigma = \sigma(\dot{\gamma})$ 
for the repulsive particles with $u=0.0$ (left)
and attractive particles with $u=0.15$ (right). The data 
are shown with symbols, while the dashed lines are 
fits to the Herschel-Bulkley form, Eq.~(\ref{hb}).
Horizontal arrows indicate the stress values 
at which constant stress simulations are 
performed, see Figs.~\ref{repstress1} and \ref{attstress1}.
Notice the very different stress scales used in both panels.
\label{flowu0u15}}
\end{figure}

In Fig.~\ref{flowu0u15}, we show the flow curves $\sigma=\sigma(\dot{\gamma})$
obtained by averaging the stress over the entire system and very long 
times in  our simulations with various values of the imposed shear rates.
For both repulsive ($u=0$) and strongly attractive ($u=0.15$) 
interactions, we find that the resulting flow curves are monotonic
functions, thus ruling out the possibility that non-monotonic flow curves
could arise in dense athermal systems with or without adhesive forces. 
Additionally, we find that all flow curves 
are well-described by an Herschel-Bulkley fitting function,
which we write as: 
\begin{equation}
\sigma_{\rm hb} = \sigma_d \left( 1 + (\tau_c \dot{\gamma} )^n \right),
\label{hb}
\end{equation}
which contains three fitting parameters. The first parameter
is the dynamic yield stress, $\sigma_d$, defined as the 
$\dot{\gamma} \to 0$ limit of the measured stress.
The second parameter, $n$, determines the shear-thinning
behaviour observed for shear rates that are large enough 
to be away from the ``yield regime'' (defined by $\sigma \approx \sigma_d$). 
The third parameter, $\tau_c$, has the dimensions of a timescale.
It indicates at which imposed shear rate the flow 
is close to the yield regime, $\tau_c \dot{\gamma} \ll 1$, 
or far from it,  $\tau_c \dot{\gamma} \gg 1$.
The obtained values of these fitting parameters
are further discussed in Sec.~\ref{rescaling}.

In the study of the formation of shear-bands in Lennard-Jones 
glass~\cite{varnik}, evidence was found of a multi-valued 
flow curve by including also the behaviour of the system
at rest, $\dot{\gamma}=0$.  
This multi-valued nature was due to the fact that the value of the static 
yield stress, $\sigma_s$, determined by checking for the onset of 
flow by applying an increasing 
external stress on a quiescent amorphous state~\cite{varnikyield}, 
was found to be higher than the dynamical yield stress $\sigma_d$
defined from finite shear rates measurements.
A strict inequality, $\sigma_s > \sigma_d$, indeed opens 
a stress window, $\sigma \in [\sigma_d, \sigma_s]$, 
where the system can either be at rest,
$\dot{\gamma}=0$, or flow at finite rate
$\dot{\gamma} > 0$, 
with the possibility that both solutions coexist in 
space, thus possibly 
giving rise to flow inhomogeneities~\cite{berthier03}.

Having ruled out non-monotonic flow curves 
at finite shear rates, we thus decided 
to check whether this scenario is a possible explanation 
of the observed flow heterogeneities. Unfortunately,
determining $\sigma_s$ numerically is not an easy 
task~\cite{varnikyield}. To proceed, we
performed simulations with an imposed constant stress, $\sigma$,
to determine whether the static and dynamic yield stresses
differ for our systems. By definition, the system should 
not flow when $\sigma < \sigma_s$. 

\begin{figure}
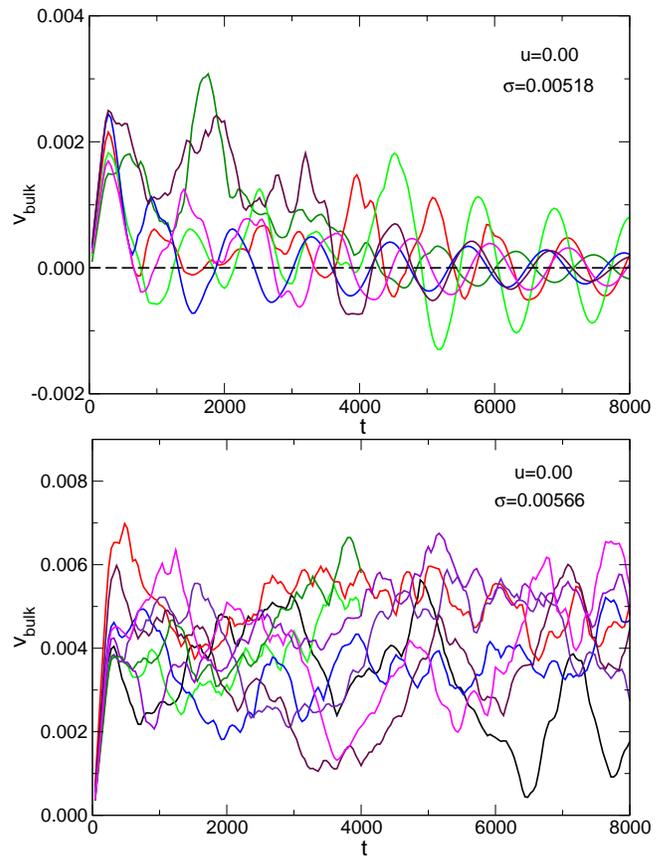

\psfig{file=fig9a.eps,width=8.5cm,clip}
\psfig{file=fig9b.eps,width=8.5cm,clip}
\caption{For repulsive particles, $u=0$,
velocity $v_{\rm bulk}$ of the center of mass of 
the system in the direction of the imposed 
shear stress starting from a configuration at rest. Different
colors correspond to independent initial conditions.
Comparison between both panels shows that the 
static yield stress lies in between the 
two simulated stress values, $\sigma=0.00518 < \sigma_s < \sigma=0.00566$.
\label{repstress1}}
\end{figure}

Using the global flow curves shown in Fig.~\ref{flowu0u15},
we select for $u=0$ two different values of stress to carry out 
our constant stress simulations. The first value
is $\sigma=0.00518$, which corresponds to $\dot{\gamma}=2.5
\times{10^{-5}}$. Second,  we use $\sigma=0.00566$, which corresponds 
to $\dot{\gamma}={10^{-4}}$. Both values are indicated with 
horizontal arrows in Fig.~\ref{flowu0u15}, and the 
results for these runs are presented in 
Fig.~\ref{repstress1}. To determine whether the system
flows or not, we simply measure the average 
velocity $v_{\rm bulk}$ of the system in the direction of the applied 
stress. As initial conditions for these runs, 
we choose arrested states corresponding to 
local minima in the energy landscape of the soft disks.

The top panel in Fig.~\ref{repstress1} 
shows that for all the different trajectories at 
$\sigma=0.00518$, the motion of the fluid particles quantified
by the average velocity $v_{\rm bulk}$ eventually stops at long times. 
On the contrary, for $\sigma=0.00566$, the flow 
continues at some finite velocity $v_{\rm bulk}$. Thus, this tells us that 
$\sigma_s$ lies in between these two stress values, and that 
indeed it is larger than the dynamic yield stress $\sigma_d$ 
for this system. The situation is therefore similar
to the observations reported for a Lennard-Jones glass~\cite{varnik}. 
The fact that there is no flow at $\sigma=0.00518$ indicates why the
heterogeneities are more significant during the simulations at 
$\dot{\gamma}=2.5\times{10^{-5}}$, in comparison to the
flow at $\dot{\gamma}={10^{-4}}$, see Fig.~\ref{repfig2}.

However, one should note that earlier numerical studies 
have reported that the difference between $\sigma_s$ 
and $\sigma_d$ seems to decrease (albeit quite 
slowly) as the system size is increased 
towards the thermodynamic limit~\cite{xu,roux}. Thus, 
this indicates that the corresponding degree of flow heterogeneties 
could also have finite size effects, despite the fact that our systems
already comprise a large number of particles, $N \approx 10^4$, 
and that the channel we use is about 100 particles
wide. We also note the following paradox. Given that 
the lengthscale of kinetic heterogeneities grows
as the shear rate decreases and is believed to diverge 
as $\dot{\gamma} \to 0$ in systems characterized by a yield 
stress~\cite{claus}, 
for any finite size system there should exist a shear rate below
which finite size effects become relevant, and the mechanism
mentioned above for the appearance of strong flow inhomogeneities
could then become relevant as well.

\begin{figure}
\psfig{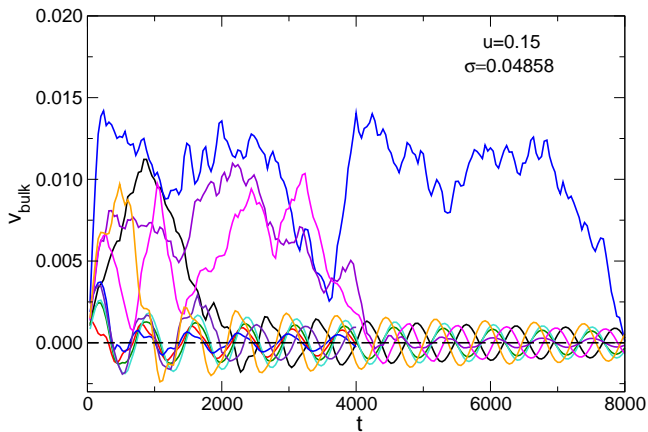}
\caption{Same as Fig.~\ref{repstress1}
for attractive particles, $u=0.15$, and a stress value
shown with a horizontal arrow in Fig.~\ref{flowu0u15}
corresponding to $\dot{\gamma}=10^{-4}$.
For all initial states, the system remains at rest, showing
that $\sigma_s > 0.04858$.
\label{attstress1}}
\end{figure}

We now switch to the attractive system with $u=0.15$ and 
ask whether forcing the system with a stress value 
corresponding to $\dot{\gamma}=10^{-4}$ induces flow 
in the system as it does for repulsive particles.
The corresponding magnitude of the external stress, 
$\sigma=0.04858$, is indicated by an arrow in 
the right panel of Fig.~\ref{flowu0u15}. The resulting data  
are plotted in Fig.~\ref{attstress1}, and should be compared
to the bottom panel of Fig.~\ref{repstress1}.
We observe that for all initial states, the system eventually comes to 
rest. Thus, in this case, the static yield stress $\sigma_s$ is
larger than the shear stress corresponding to $\dot{\gamma}=10^{-4}$.
This result is in agreement with our earlier
observation that at this shear rate, attractive systems
display much stronger flow inhomogeneities than repulsive ones,
recall Fig.~\ref{momdat1e4}.

Therefore, we conclude from this section that
the existence of distinct values for static and dynamic
yield stresses accounts well for flow inhomogeneities in 
our system, as opposed to a non-monotonic flow curve at finite shear rates.
Moreover, we also showed that the stress window where this competition
becomes relevant corresponds to shear rates values that become larger
when attraction is increased, as shown by our constant stress
simulations. We finally remark that these latter results 
suggest that by making simulations at constant 
shear stress, the fluid velocity would remain zero 
as long as $\sigma < \sigma_s$, and would jump discontinuously
to a finite value at $\sigma_s$.
This is nothing but the ``viscosity bifurcation''  
observed in several experiments~\cite{coussot4,dacruz,ragouillaux}.
In this language, our results imply that increasing the attraction
increases the value of the ``critical'' shear rate $\dot{\gamma}_c$
at which a steady state flow appears, which seems consistent
with experimental results~\cite{paredes}.

Thus, we come to the conclusion that the effect of particle
adhesion is to continuously increase the critical shear rate
above which stable linear profiles exist. 
Therefore, for adhesive particles,
the regime below the critical shear rate becomes easily accessible in
numerical simulations, leading to observation of more prominent
inhomogeneities in flow. However, we do not obtain numerical evidence 
that some novel physics comes in, 
such as for instance a slower restructuration process, as advocated 
in Refs.~\cite{coussotovarlez,vincent,kirsten2}). 

We now seek a more microscopic explanation of these effects.

\subsection{``Universal'' rescaling of flow curves}
\label{rescaling}

\begin{figure}
\psfig{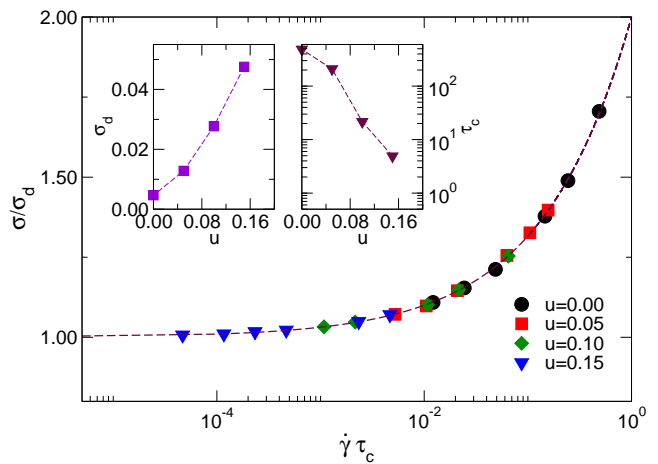}
\caption{Scaled representation of the flow curves for all 
parameters, using the scaled variables in Eq.~(\ref{scaled}).
The dashed line is the simple functional form in Eq.~(\ref{functional}).
Inset: Variation of the dynamic yield stress $\sigma_d$ (left),
and of the timescale $\tau_c$ (right) with the strength of the attractive 
forces $u$.
\label{flowcurv2}}
\end{figure}

We thus see the emergence of a universal scenario for the existence of 
flow heterogeneities, irrespective of the interparticle interactions.
Firstly, we observed that for both attractive or repulsive interactions,
the  heterogeneous flow corresponds to the regime bounded by values of 
static and dynamic yield stresses.
Moreover, the flow curves for all the different systems ($u=0.0, 0.05, 0.10, 
0.15$) are well fitted with the same
functional form, Eq.~(\ref{hb}). 
We find empirically that the shear-thinning exponent 
$n$ varies very little around the value $n \approx 0.5$ with no systematic
trend. Thus, we fix $n= \frac{1}{2}$ in the following and determine
$\sigma_d$ and $\tau_c$ from a fit to the data.
This analysis thus suggests that the flow curves 
for all our systems can be collapsed by 
using the scaled variables
\begin{equation}
y = \frac{\sigma}{\sigma_d}; \quad 
x = \tau_c \dot{\gamma};
\label{scaled}
\end{equation}
onto the simple functional form: 
\begin{equation}
y = 1 + \sqrt{x}.
\label{functional}
\end{equation}
This procedure is applied in Fig.~\ref{flowcurv2}, 
where the two insets display the evolution with the strength
of the attraction of the two parameters 
$\sigma_d$ and $\tau_c$ obtained by fitting the flow curves.

We find that the yield stress $\sigma_d$ increases with 
the attraction $u$, which is expected as the attraction 
should indeed make the emulsion more cohesive and harder to
deform. A more surprising result is that 
the timescale $\tau_c$ is found to decrease dramatically 
by about two orders of magnitude between $u=0$ and 
$u=0.15$. We shall dwell on this unexpected result
in the following section.

The significant observation following the data collapse 
shown in Fig.~\ref{flowcurv2} is that, for the 
range of imposed shear rates explored in our simulations, 
the data points corresponding to increasing 
attraction correspond to dramatically decreasing values of the scaled variable 
$x=\tau_c \dot{\gamma}$ in the master curve.
Thus, varying the degree of attraction allows us to ``slide'' along this 
master curve, so that the most attractive system is 
effectively much closer to the yield stress regime, 
where strongly heterogeneous flow can be expected. 
On the other hand, repulsive particles correspond to a more 
fluidized segment of the flow curve, which explains why 
only relatively short-lived inhomogeneities are observed in 
that case. 

While framing their model for shear-banding
in Ref.~\cite{coussotovarlez}, 
the authors obtained a similar rescaling
of experimental data for varying degrees of loaded emulsions. However, 
they attribute an increasing tendency towards shear-banding to 
an increasing timescale for ``restructuration'' of the material
under shear with higher degree of loading (stickiness), 
which is a trend opposite to what we observe here.
For clarity, we should emphasize again 
that at the large value of the volume fraction ($\phi=1.0$) 
that we have studied here, the material is not a gel but a highly 
homogeneous material.
For this reason, we believe that 
local fluctuations in the volume fraction are not significant, 
which also rules out the flow-concentration coupling scenario proposed
in another model~\cite{besseling} for our jammed systems.

The above analysis implies that 
varying the attraction changes the nature 
of inhomogeneities by affecting the scaled 
variable $x=\tau_c \dot{\gamma}$ through the evolution of 
$\tau_c$. Clearly, the same result would be obtained 
by varying instead the shear rate for a given value 
of $\tau_c$. This reasoning is consistent 
with the data shown in Fig.~\ref{repfig2} which established that
inhomogeneities increase when the shear rate decreases.

\subsection{Microscopic interpretation of  timescale $\tau_c$}

The final piece of information that we need to gather
is a microscopic understanding of the strong variation
with the strength of attractive forces 
of the intrinsic timescale $\tau_c$ appearing in the 
global flow curves, Eq.~(\ref{hb}). Under
an applied shear, the system constantly renews its structure.
We believe that $\tau_c$ quantifies the timescale needed
for the local stress relaxation to occur.
Thus, when the typical timescale $1/\dot{\gamma}$
associated with the imposed shear flow is much slower than
$\tau_c$, i.e. when $\tau_c \dot{\gamma} \ll 1$, 
the system is effectively in the near quasi-static
regime, and $\sigma \approx \sigma_d$. In the opposite
regime, $\tau_c \dot{\gamma} \gg 1$, the system 
has not enough time to relax the stress whose averaged value
increases above the yield stress. 

In this view, 
$\tau_c$ is a timescale associated to relaxation of the stress after 
deformation. To confirm this interpretation, we
conduct the following numerical experiment. 
For the different values of attraction from $u=0$
to $u=0.15$, we apply an instantaneous external strain
in the $x$-direction of magnitude $\epsilon$, starting
from an initially relaxed state 
(which corresponds to a local minima in the energy landscape).
We then allow this strained configuration to relax~\cite{hatano,zausch}.
To fix the value of $\epsilon$, we seek a compromise between 
a very small value where only trivial elastic deformation would occur,
and a very large value which would amount to starting 
from a fully random configuration. We have studied two values, 
$\epsilon=0.1$ and $\epsilon=0.2$, because they typically correspond 
to the magnitude of the strain where plastic deformations 
occur in quasi-static simulations~\cite{claus2}. We find 
similar results in both cases and show results
obtained for $\epsilon=0.2$.

\begin{figure}
\psfig{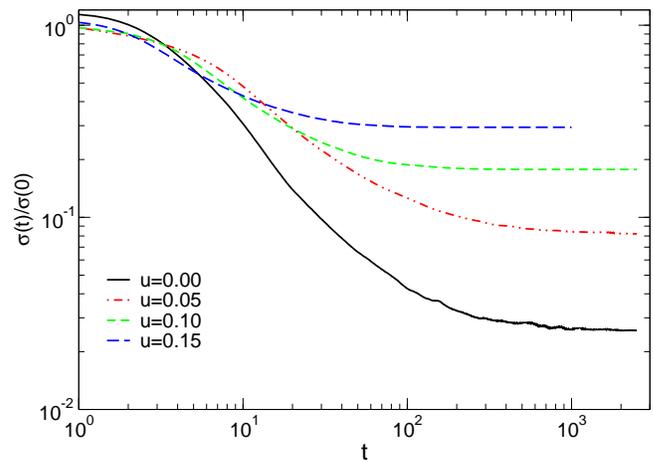}
\caption{Relaxation of the stress after a step strain of amplitude 
$\epsilon=0.2$ for the different strengths of interaction $u$.
The amount of relaxed stress and the timescale for 
relaxation both decrease strongly with increasing the attraction $u$.}
\label{stressrelax}
\end{figure}

We present our data for stress relaxation after 
these step strains in Fig.~\ref{stressrelax}, 
where each curve is averaged over several (typically
10) initial states. To ease the
comparison betwen different values of the attraction,
we plot $\sigma(t) / \sigma(0)$, where
$\sigma(0)$ is the stress value right after the step strain 
has been imposed.

Several remarks can be made based on this figure. Altogether
the stress exhibits a first rapid decay, followed by a plateau at long time
scale. First,
the short timescale for the stress relaxation
is observed to decrease strongly with increasing the attraction strength. 
This is 
in line with the result for the time-scale $\tau_c$ in Fig.\ref{flowcurv2}.  
Second,
 it is evident that the degree of stickiness influences the amount 
of stress that is relaxed after the step strain, which 
becomes smaller for stronger adhesion:  stickiness increases the amount
of residual stress that the system can store.
A possible interpretation is that after deformation
adhesive particles very rapidly ``stick'' to the neighboring
particles to minimize locally the potential energy, and 
remain subsequently in this local minimum, while 
more collective, slower moves relaxing larger amounts of stress,
are more likely to occur for purely repulsive particles. 

In the present system, shear localization
is enhanced by adhesion because attractive forces
accelerate the timescale for restructuration after deformation,
thus effectively moving the system closer to the quasi-static,
yield stress regime where heterogeneities are more pronounced.
Furthermore the increase of residual stresses for 
adhesive particles
does stabilize the transient shear-bands over longer times.

\section{Conclusion}

\label{conclusion}

We carried out simulations of two-dimensional jammed systems to 
study the nature of shear flow heterogeneities and the influence 
of the degree of attractive interaction between particles. 
We demonstrated that, independent of the
nature of interactions, the flow behaviour could be described by a 
universal flow curve, with increasing attraction
resulting in a flow which is more and more influenced by the 
proximity to the yield stress regime. By entering a regime 
of stress values bounded by the static and dynamic 
yield stress, shear localization is strongly promoted, 
and with increasing attraction,
one observes enhanced and very long-lived flow 
inhomogeneities. For the most strongly 
adhesive particles, we find that velocity 
profiles do not become linear even when averaged 
over very large deformations, suggesting that 
in these systems a linear flow profile is 
actually unstable. However, these inhomogeneities 
do not take the form of simple, permanent shear-bands.

Our results are reminiscent of long-lived 
flow inhomogeneities measured experimentally 
in simple yield stress fluids~\cite{manneville}. 
Interestingly, similar to our adhesive soft system, 
carbopol is measured to exhibit a monotonic 
Herschel-Bulkley flow-curve \cite{manneville}, 
and does not exhibit thixotropy, nor aging. An important difference 
is that the experiments report long transients when the shear is turned on,
while we worked here in steady state.
On the theoretical side, similar conclusions about 
the transient nature of shear-bands were
reached in Refs.~\cite{manning,Moorcroft,roux2} on the basis of 
spatially resolved coarse-grained models
using the idea that after long aging the static yield stress value 
can become appreciably larger than the dynamic yield stress.
We emphasize that, by contrast to these works, the distinct values for these 
two yield stresses we report here do not depend on the aging time
since our systems are fully athermal. 

By contrast, fully permanent flow localization 
taking the form of simple shear-bands has been predicted through 
several mesoscopic 
models~\cite{vincent,jagla,coussotovarlez,kirsten2,fielding}, 
which all attempt to describe 
a self-consistent coupling of the yielding mechanism to 
the structural reorganisation 
({\it e.g.}, softening mechanisms,  timescales separation in the 
structural relaxation, etc.).
While these descriptions lead to the occurence of permanent 
bands, this flow behavior is
systematically associated with the appearance of a 
non-monotonic behavior of the flow curve $\sigma(\dot\gamma)$
at finite shear rate.  Although experimental results, 
simulations and the various theoretical models all suggest
that this non-monotonicity is a necessary condition to observe
permanent shear-bands, we demonstrate here that 
very long-lived, strongly non-linear flow profiles
can be observed without it.

Clearly, to make progress and to go beyond the above observations
and conflicting predictions, 
it would be desirable to make explicit connections
between the microscopic physical parameters in experiments 
or simulations--such as the adhesive forces considered 
in this work--, and the mesoscopic phenomenological quantities 
introduced in the various simplified theoretical 
models, such as for instance the distinct timescales which 
enter the definition of coarse-grained
elasto-plastic models~\cite{vincent,coussotovarlez,kirsten2}, or 
the effective `noise' temperature in the SGR models and its 
numerous variants~\cite{fielding,Moorcroft}.

Altogether, our results suggest that ``shear localization''
actually denotes a broad variety of physical 
behaviours, and further experiments 
performed with controlled 
model systems with tunable adhesion would be required to 
investigate the intimate connection between 
structural recovery and shear localization
along the lines of the present work. Similarly, 
it would be interesting to design a simple microsopic model
yielding the type of permanent shear-bands envisioned 
by coarse-grained models, for instance by studying 
simple atomistic models for colloidal gels.

\acknowledgments

We thank ANR SYSCOM for financial support and 
C. Barentin,
J.-L. Barrat,
A. Colin, 
T. Divoux, 
C. Heussinger,
K. Martens 
for useful discussions.

\end{document}